# Near-coherent quantum emitters in hexagonal boron nitride with discrete polarization axes

Jake Horder[1], Dominic Scognamiglio[1], Ádám Ganyecz[2,3], Viktor Ivády[3,4,5], Nathan Coste[1], Mehran Kianinia[1,6], Milos Toth[1,6] and Igor Aharonovich[1,6]

1. School of Mathematical and Physical Sciences, University of Technology Sydney, Ultimo, New South Wales 2007, Australia

2. Wigner Research Centre for Physics, PO Box 49, H-1525, Budapest, Hungary

3. MTA–ELTE Lendület "Momentum" NewQubit Research Group, Pázmány Péter, Sétány 1/A, 1117 Budapest, Hungary

4. Department of Physics of Complex Systems, Eötvös Loránd University, Egyetem tér 1-3, H-1053 Budapest, Hungary

5. Department of Physics, Chemistry and Biology, Linköping University, SE-581 83 Linköping, Sweden

6. ARC Centre of Excellence for Transformative Meta-Optical Systems, University of Technology Sydney, Ultimo, New South Wales 2007, Australia

milos.toth@uts.edu.au; igor.aharonovich@uts.edu.au

**Abstract**

Hexagonal boron nitride (hBN) has recently gained attention as a solid state host of quantum emitters. In particular, the recently discovered B centers show promise for integration in scalable quantum technologies thanks to site-specific defect generation and reproducible wavelength. Here we employ spectral hole burning spectroscopy and resonant polarization measurements to observe nearly-coherent hBN quantum emitters, both as singles and in ensembles, with three discrete polarization axes indicative of a $C_{2v}$ symmetry defect. Our results constitute an important milestone towards the implementation of hBN quantum emitters in integrated quantum photonics.

1. Introduction

Single photon sources are a central component of scalable photonic quantum technologies [1,2]. A key prerequisite needed for scalable quantum photonic applications is to have coherent single photon sources. In solid state systems, this is often challenging as the spectral lines are broadened by either spectral diffusion or by the presence of phonons. Many solid state systems have been investigated as deterministic coherent sources of quantum light, including semiconductor quantum dots [3,4], color centers in three dimensional bulk crystals [5–7], and two dimensional (2D) van der Waals crystals [8]. The 2D material hexagonal boron nitride (hBN) has attracted increasing interest due to reports of bright, stable linearly-polarized quantum emitters with a broad range of emission wavelengths [9–13].

Of the various quantum emitters observed in hBN, the B center defect is compelling because it can be engineered on demand with a site-specific fabrication technique and has a highly-reproducible emission wavelength around 436 nm [14–16]. Such reproducibility is a crucial feature for future integration of these emitters into scalable devices for photonic quantum technologies. Consequently, the defect was used in preliminary demonstrations of photon indistinguishability [17] and of incorporation in rudimentary devices [18–20]. However, exploitation of the B center in practical quantum coherent systems requires the combination of a long photon coherence time, and a discrete number of well-defined polarization axes within the hBN crystal lattice.

Here we use spectral hole burning (SHB) spectroscopy to observe near-lifetime-limited lines as narrow as 150 MHz, in both an isolated B center defect, and in individual defects within an ensemble. The SHB technique allows to sidestep linewidth broadening due to spectral diffusion [21] induced by carbon-doping of the hBN sample, which is often necessary for the generation of B centers [22]. We also perform resonant polarization measurements to show that the B center emission has one of three discrete in-plane polarization axes separated by 60°. We discuss these results within the framework of the Jahn-Teller (JT) distortion of a split interstitial defect, yielding a $C_{2v}$ symmetry within the hBN crystal lattice.

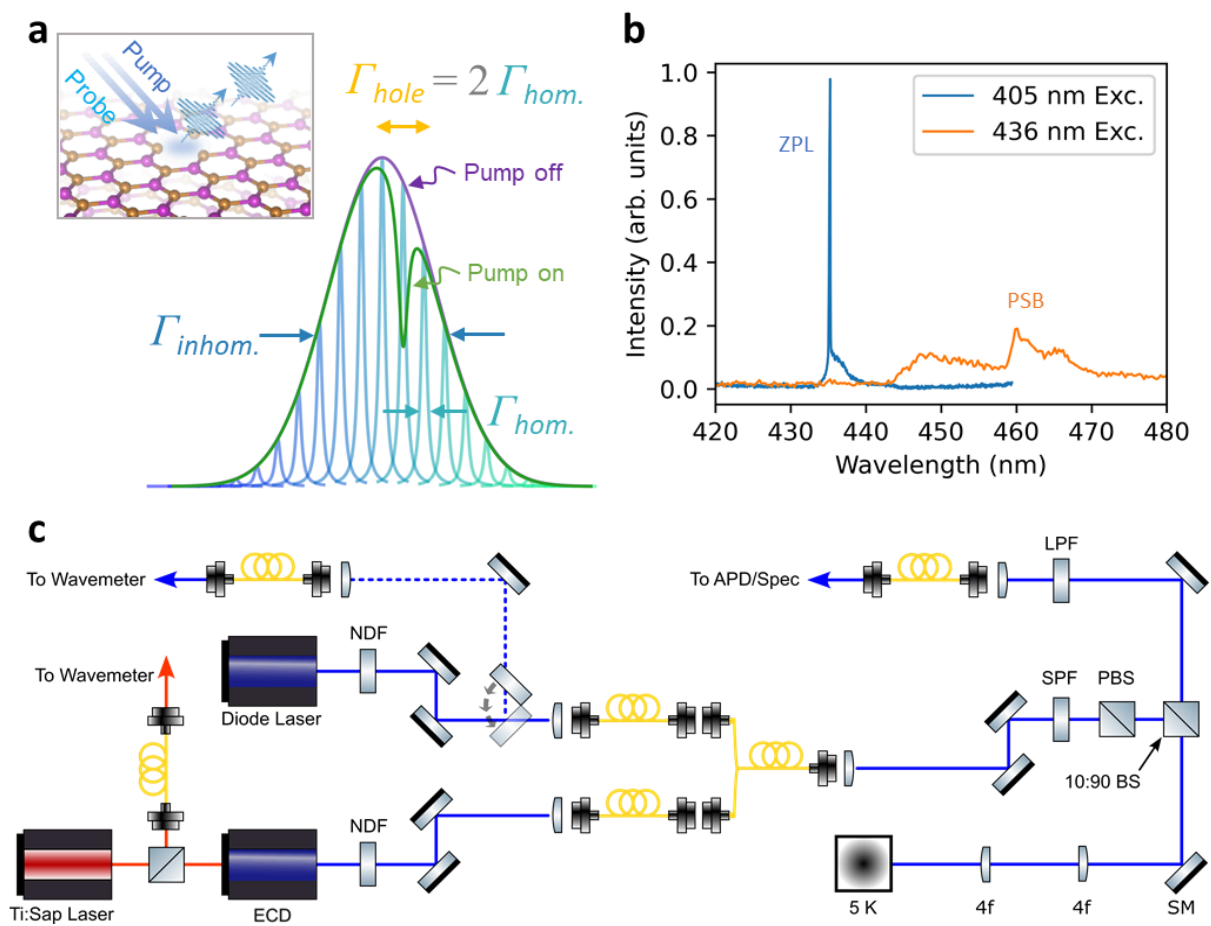

*Figure 1. B center characterisation and optical setup. a) Schematic of spectral hole detected by scanning a probe laser over an inhomogeneously broadened spectrum of width $\Gamma_{inhom.}$, where a stationary pump laser creates a dip of width $\Gamma_{hole}$ equivalent to twice the homogeneous linewidth $\Gamma_{hom.}$. Inset: schematic representation of single photon generation vai pump-probe resonant excitation of a defect state in the hBN lattice. b) The zero phonon line*

*(ZPL) of the B center at 436 nm is evident under non-resonant excitation (405 nm Exc.). When exciting resonantly (436 nm Exc.) the excitation laser is filtered from the collected emission using a 442 LPF, leaving only the phonon sideband (PSB). c) Optical setup for two-laser resonant excitation. Separate lasers are coupled to a single launching point using a fiber coupler. Ti:Sap laser (M Squared) linewidth 50 kHz, diode laser (Toptica) linewidth 300 kHz; External cavity doubler (ECD); Neutral density filter (NDF); short pass filter (SPF); polarizing beam splitter (PBS); 10:90 (R:T) beam splitter (BS); scanning mirror (SM); 4f lens (4f); long pass filter (LPF).*

## 2. Materials and Methods

A schematic illustration of the SHB experiment and the B center is shown in Figure 1a. Whilst the exact atomic structure of the defect is still a matter of debate, two promising candidates are the split carbon and split nitrogen configurations [16,23]. To fabricate the B centers, a Si/SiO2 (285 nm) substrate was first cleaned by sonication in acetone for 30 min, then rinsed with isopropanol and dried using nitrogen before exposure in UV ozone bath for 10 min. hBN flakes were mechanically exfoliated on the substrate using scotch tape, then the sample was annealed on a hotplate for 1 hr at 500ºC. Electron beam irradiation was used to create a 5x5 array of B center ensembles at 5 μm spacing using a dual beam microscope with 0.8 nA current and 5 keV beam energy.

The emission spectrum of a single B center is shown in Figure 1b. It consists of a prominent zero phonon line (ZPL) at 436 nm, and a phonon sideband (PSB) at 460 nm. Unless stated otherwise, all measurements are done on resonance at 5 K. All data are collected from a home built confocal microscopy setup with 0.82 NA objective lens (Attocube). Resonant excitation is performed with a frequency doubled Ti:sapphire scanning laser (M Squared) and External Cavity Diode Laser (ECDL, Toptica), and photons collected from the PSB using a 442 nm long pass filter are directed to a spectrometer (Andor) or single photon counters (Excelitas) through multimode fiber. PLE data in hole burning measurements are collected over 25 ms integration time, and the scanning (probe) laser wavelength is scanned at a rate of 1 GHz/s. The full linewidth is measured by manually incrementing the probe laser wavelength, and averaging the APD count rate over 10 s acquisition at each point.

SHB measurements are implemented by saturating a transition subspace using a stationary, high power resonant laser (pump laser) and scanning across the whole transition with a secondary resonant laser (probe laser). Figure 1c shows the schematic configuration of our setup, where the two resonant lasers were made to excite in the same mode by first coupling to one single mode launching fiber via a fiber coupler, and then passing the excitation path through a polarizing beam splitter. SHB spectroscopy was employed as it enables direct measurement of the phonon-broadened linewidth, and hence the best case coherence time in the absence of spectral diffusion [24–30].

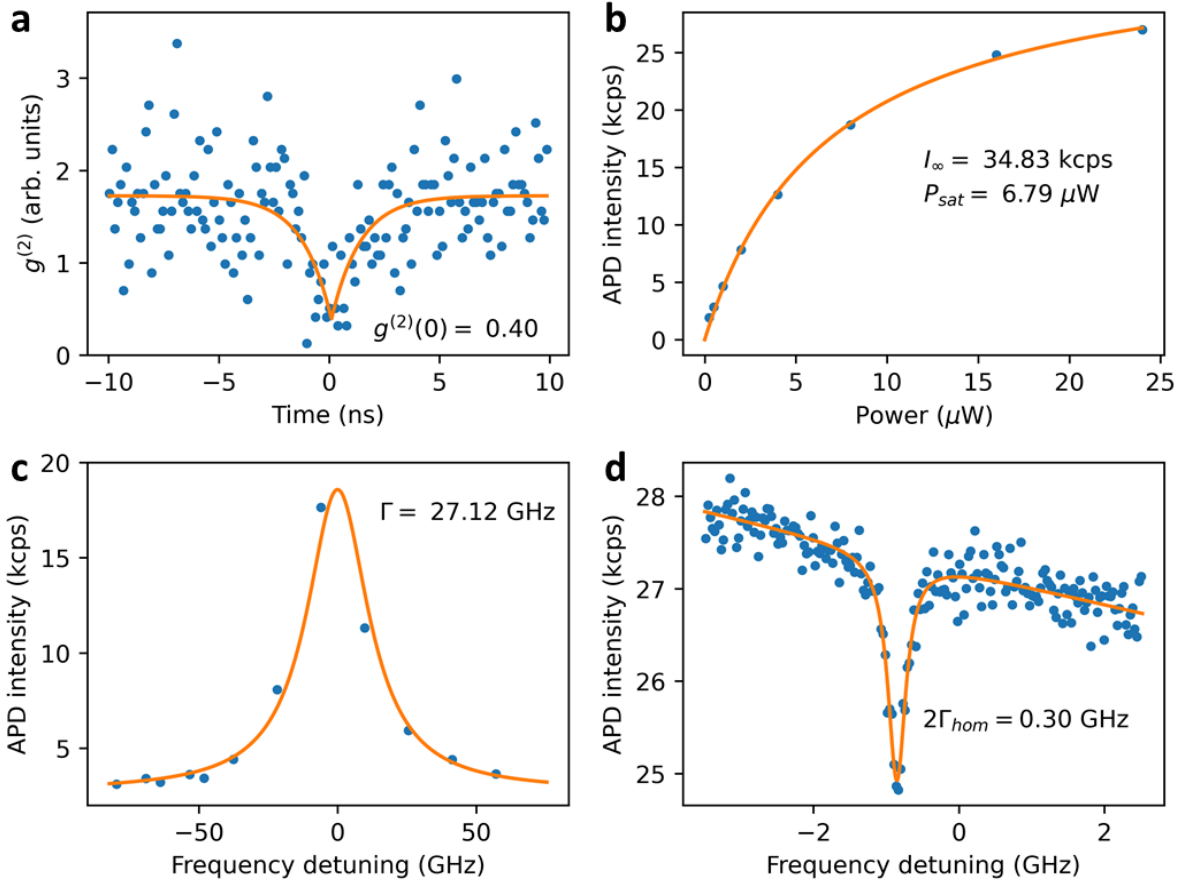

*Figure 2. Single emitter spectral hole burning. a) Autocorrelation measurement using 6.8 μW of resonant laser excitation. Fitting with a two level model yields a minimum of 0.40 at zero time delay. b) Power saturation trend under resonant excitation. The fit yields a maximum intensity of 34.83 kcps, with a half-maximum intensity at 6.79 μW. c) Full resonant photoluminescence spectrum performed with 6.8 μW of probe laser only. A Lorentzian fit results in a FWHM of 27.12 GHz, and the relative frequency is defined with respect to the center of the fit. d) A hole is burned in the spectrum using an additional 10 μW stationary pump laser. The Lorentzian fit yields a hole width of 0.30 GHz.*

3. **Results and Discussions**

A resonant autocorrelation measurement using 6.8 μW of laser power from a single B center is shown in Figure 2a, with a fit to the data yielding a purity of $g^{(2)}(0) = 0.40$. The bunching above unity at small times is likely due to excitation at the saturation power, which was found to be 6.79 μW as is shown in Figure 2b. The relatively low purity is attributed to a poor signal-to-noise ratio.

A scan over the entire resonant spectrum at saturation power is shown in Figure 2c, where the frequency detuning is relative to the ZPL at approximately 687 THz. Notably, the observed linewidth of 27 GHz is an order of magnitude broader than typical linewidths that were reported previously [15,16,31]. This is indicative of substantial spectral diffusion, which is expected as in the present work we employed heavily C-doped hBN that contains a relatively high density of charge traps.

To implement SHB, a second 10 μW resonant pump laser was employed. Under two-laser resonant excitation, a narrowband hole was burned with a linewidth of ~300 MHz, as shown in Figure 2d. The hole width corresponds to twice the homogeneous linewidth of the emitter in the absence of spectral diffusion [21], i.e. subject only to phonon-induced dephasing and power broadening. Reported B center lifetimes of 1.8-2.6 ns designate a lower bound for the linewidth of ~60-90 MHz [31,32], and our measured homogeneous linewidth of 150 MHz represents a coherence time of $T_2 = 1/(\pi\Gamma_{hom}) = 2.12$ ns, or $T_2 \approx T_1$, indicating single photon coherence close to the maximum lifetime-limited coherence. We note that this is the narrowest linewidth reported from a stable hBN emitter to date.

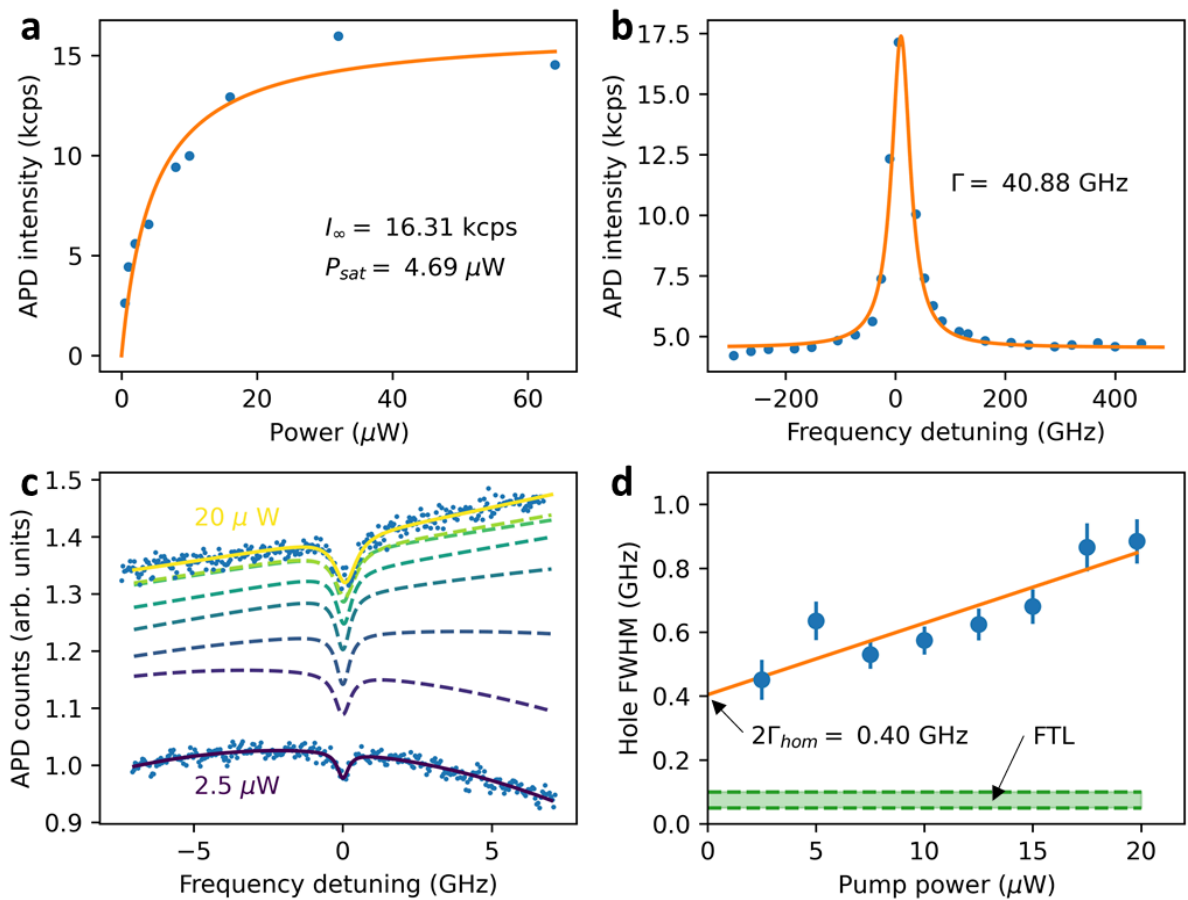

Figure 3. Power dependent spectral hole burning. a) Power saturation trend under resonant excitation. The fit yields a maximum intensity of 16.31 kcps, with the half-maximum intensity observed at 4.69 μW. b) Full resonant photoluminescence spectrum performed with 10 μW of probe laser only. A Lorentzian fit results in a FWHM of 40.88 GHz, and the relative frequency is defined with respect to the hole burning wavelength. c) A pump laser at increasing powers is used to burn a hole in the peak of the spectrum. The data and fitted curves for the lowest (2.5 μW) and highest (20 μW) pump powers are shown, while only the fits for the intermediate powers are shown for clarity. All data sets are normalized with respect to the average intensity of the lowest power data set. d) The spectral hole widths extracted from the fits in (c) as a function of pump power. A linear fit yields a minimum hole

*width of 0.40 GHz. The green shaded region represents the Fourier transform limit (FTL) linewidth range.*

To study the homogeneous broadening of the B centers, we repeat the SHB on an ensemble of emitters. An ensemble site was identified from a resonant confocal scan, with the power saturation behavior depicted in Figure 3a showing an intensity of 16.31 kcps and a saturation power of 4.69 μW. The coarse resonant scan in Figure 3b shows a broad linewidth of ~40 GHz, which we attribute to spectral diffusion (in addition to power broadening) [27]. The pump power affects the width of holes burned into spectra [26,30]. Hence, we varied the pump laser power from 2.5 μW up to 20 μW, keeping the pump wavelength fixed at the ZPL wavelength. Figure 3c shows the holes revealed by scanning the probe laser over the pump wavelength, where only the fits are plotted for the intermediate powers for clarity. All data sets are normalized to the average intensity seen when the pump laser is set to 2.5 μW, and the diminishing increase in intensity of the higher-power fitted curves reflects the plateau in intensity seen in the power saturation data in Figure 3a.

The width of the Lorentzian hole in each spectrum is plotted as a function of pump laser power in Figure 3d. The FWHM shows a clear power dependence, and a linear fit allows for an estimate of ~400 MHz for the hole width in the absence of power broadening. From this value we obtain a power-independent homogeneously broadened linewidth of ~200 MHz for this B center ensemble, indicating a nearly lifetime-limited line with the moderate degree of broadening likely due to electron-phonon coupling and the JT effect, as discussed below.

Remarkably, the emitters' linewidth in the SHB experiments from the ensemble is on par with the linewidth obtained from a single site, directly hinting at the uniform nature of the emitters. This represents a major advance for the future implementation of the B center in quantum applications requiring indistinguishable photons from multiple sources.

To complete the spectroscopic analysis of the B centers, we study their emission polarization properties employing resonant excitation. First, we identified B centers within close proximity in a region of the hBN flake free of grain boundaries. A quarter wave plate was used to transform the linear excitation into circularly polarized excitation so that emitters with all possible absorption dipole orientations were equally likely to be visible in the confocal scan. Emission polarization was measured for each B center site, and a sinusoidal fit to each data set is shown in Figure 4a. The fits have the form $p = \kappa \sin(2\theta - \varphi) + \beta$, where $\kappa$ is the sinusoidal amplitude, $\theta$ is the in-plane rotational angle, $\varphi$ is a phase offset accounting for orientation, and $\beta$ is a vertical offset. The polarization visibility is defined as $V = (p_{max} - p_{min})/(p_{max} + p_{min})$. All data sets are found to have a polarization visibility greater than 0.6, indicating linearly polarized emission typical of an in-plane optical transition dipole.

Most strikingly, the orientations among the polarization plots show a strong clustering at three distinct angles, with 60° separation between clusters. In Figure 4b, the phase parameter $\varphi$ from each polarization fit is plotted, in order of increasing value. There are three distinct groups evident, corresponding to three preferential angles of emission orientation, most likely aligned with the crystal directions of hBN.

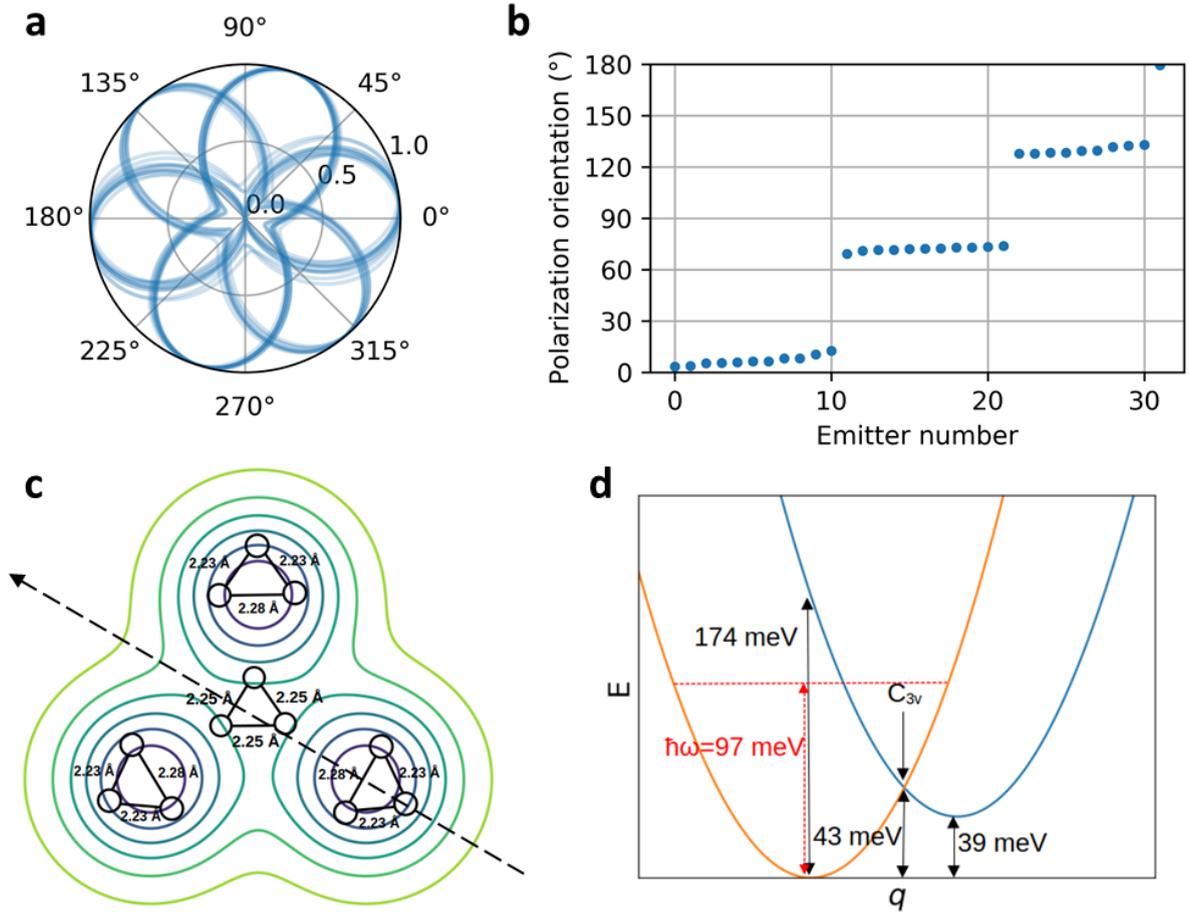

*Figure 4. Emission polarization statistics. a) Emission polarization for several dozen emitters under 10 µW of resonant excitation, fit with a sinusoidal function. b) Orientation angle extracted from the fit of each polarization plot in (a), sorted by increasing angle. c) Potential energy surface of the excited state of the nitrogen split interstitial defect. Colored curves represent low (blue) to high (green) iso-energy curves on the potential energy surface. Triangles with numbers indicate the relative distance of the first neighbor boron atoms in the nitrogen split interstitial defect at the corresponding position of the potential energy surface. κd) Parameter of the JT system. The parabolas schematically depict the intersection of the potential energy surface along the dashed line in (c). The JT energy is 43 meV, while the barrier energy is 39 meV. The energy of the phonon that couples to the JT distortion in the excited state is 97 meV.*

To understand these results, we consider the proposed negatively charged nitrogen split interstitial defect, which has been suggested as a working model for the B center [23]. The defect exhibits a JT unstable excited state that goes through a spontaneous structural distortion with three possible final configurations as an outcome. Each of these configurations possesses low symmetry ($C_{2v}$) and exhibits a well-defined in-plane polarization axis. Parallel to this axis the photoluminescence intensity is maximal, while perpendicular it vanishes. This feature is well characterized by the angle-dependent intensity seen in our polarization measurements. The three possible JT distortions give rise to three different polarization axes with 120° relative orientations, which can explain the 60° rotation between planes of photon polarization seen in Figure 4b. From the analysis of the potential energy surface of the excited state shown in Figure 4c, we conclude that the system is partially dynamic JT, meaning that transition between the three distorted configurations is possible. This would induce a change in the polarization axis of the defect with time. Figure 4d provides the parameters of the JT system. The dynamic JT effect can also explain why the observed linewidths do not reach the lifetime-limited linewidths, as was also observed for other color centers [33]. Finally, we note that we cannot exclude the possibility that the B center structure could be the carbon dimer defect ($C_BC_N$), that was originally proposed as a source of the 4.1 eV emission. While on the one hand, our polarization data would support this assumption, our resonant excitation measurements, and those of other groups, suggest the B center has a high symmetry, with no permanent dipole moments. This is analogous to the more conventional inversion symmetry present in group IV defects in diamond, that result in nearly coherent emitters, without constant perturbation of the linewidths that appear in dipole-like emitters (e.g. NV center in diamond [34,35]).

## 4. Conclusions

In summary, we have used two-laser resonant excitation to implement spectral hole burning in B centers, revealing near lifetime-limited linewidths as narrow as 150 MHz for individual emitters and ensembles. Absent spectral diffusion, the B center can be expected to source coherent single photons suitable for quantum photonics applications. The uniformity of emission extends to polarization, with only three possible orientations seen across a single flake domain region. This feature could be leveraged for optimal cavity coupling, opening up emission rate enhancement and further improving the coherence properties. Together with site-specific generation of such defects with nanometric precision and reproducible wavelength, these results pave the way towards applications in integrated quantum photonics such as on-chip boson sampling [36] and photonic quantum computing [37], superradiant lasers [38] and exploring multi-emitters cavity-QED regimes [39,40].

**Funding**

Australian Research Council (CE200100010, FT220100053), the Office of Naval Research Global (N62909-22-1-2028). This research is supported by an Australian Government Research Training Program Scholarship. VI was supported by the National Research, Development, and Innovation Office of Hungary within the Quantum Information National Laboratory of Hungary (Grant No. 2022-2.1.1-NL-2022-00004) and within grant FK 145395. V.I. also appreciates the support from the Knut and Alice Wallenberg Foundation through WBSQD2 project (Grant No.\ 2018.0071).

**Disclosures**

The authors declare no conflicts of interest.

Technologies, Nat. Photonics **14**, 5 (2020).